\documentclass[]{spie}
\usepackage[dvips]{graphicx}
\usepackage{epsfig}

\def\eps@scaling{.95}
\def\epsscale#1{\gdef\eps@scaling{#1}}

\def\plottwo#1#2#3#4{\centering \leavevmode
    \epsfxsize=#2\columnwidth \epsfbox{#1} \hfil
    \epsfxsize=#4\columnwidth \epsfbox{#3}}

\def\aj{\ref@jnl{AJ}}                   
\def\araa{\ref@jnl{ARA\&A}}             
\def\apj{\ref@jnl{ApJ}}                 
\def\apjl{\ref@jnl{ApJ}}                
\def\apjs{\ref@jnl{ApJS}}               
\def\ao{\ref@jnl{Appl.~Opt.}}           
\def\apss{\ref@jnl{Ap\&SS}}             
\def\aap{\ref@jnl{A\&A}}                
\def\aapr{\ref@jnl{A\&A~Rev.}}          
\def\aaps{\ref@jnl{A\&AS}}              
\def\azh{\ref@jnl{AZh}}                 
\def\baas{\ref@jnl{BAAS}}               
\def\jrasc{\ref@jnl{JRASC}}             
\def\memras{\ref@jnl{MmRAS}}            
\def\mnras{\ref@jnl{MNRAS}}             
\def\pra{\ref@jnl{Phys.~Rev.~A}}        
\def\prb{\ref@jnl{Phys.~Rev.~B}}        
\def\prc{\ref@jnl{Phys.~Rev.~C}}        
\def\prd{\ref@jnl{Phys.~Rev.~D}}        
\def\pre{\ref@jnl{Phys.~Rev.~E}}        
\def\prl{\ref@jnl{Phys.~Rev.~Lett.}}    
\def\pasp{\ref@jnl{PASP}}               
\def\pasj{\ref@jnl{PASJ}}               
\def\qjras{\ref@jnl{QJRAS}}             
\def\skytel{\ref@jnl{S\&T}}             
\def\solphys{\ref@jnl{Sol.~Phys.}}      
\def\sovast{\ref@jnl{Soviet~Ast.}}      
\def\ssr{\ref@jnl{Space~Sci.~Rev.}}     
\def\zap{\ref@jnl{ZAp}}                 
\def\nat{\ref@jnl{Nature}}              
\def\iaucirc{\ref@jnl{IAU~Circ.}}       
\def\aplett{\ref@jnl{Astrophys.~Lett.}} 
\def\apspr{\ref@jnl{Astrophys.~Space~Phys.~Res.}}
\def\bain{\ref@jnl{Bull.~Astron.~Inst.~Netherlands}}
\def\fcp{\ref@jnl{Fund.~Cosmic~Phys.}}  
\def\gca{\ref@jnl{Geochim.~Cosmochim.~Acta}}   
\def\grl{\ref@jnl{Geophys.~Res.~Lett.}} 
\def\jcp{\ref@jnl{J.~Chem.~Phys.}}      
\def\jgr{\ref@jnl{J.~Geophys.~Res.}}    
\def\jqsrt{\ref@jnl{J.~Quant.~Spec.~Radiat.~Transf.}} 
\def\memsai{\ref@jnl{Mem.~Soc.~Astron.~Italiana}}
\def\nphysa{\ref@jnl{Nucl.~Phys.~A}}   
\def\physrep{\ref@jnl{Phys.~Rep.}}   
\def\physscr{\ref@jnl{Phys.~Scr}}   
\def\planss{\ref@jnl{Planet.~Space~Sci.}}   
\def\procspie{\ref@jnl{Proc.~SPIE}}   

\def\lesssim{\mathrel{\hbox{\rlap{\hbox{\lower4pt\hbox{$\sim$}}}\hbox{$<$}}}}
\def\gtrsim{\mathrel{\hbox{\rlap{\hbox{\lower4pt\hbox{$\sim$}}}\hbox{$>$}}}}

\def\arcsec{\hbox{$^{\prime\prime}$}}

\def\farcm{\hbox{.\kern -0.7ex\raisebox{.9ex}{\scriptsize$\prime$}}}
\def\farcs{\hbox{\kern 0.13ex.\kern -0.95ex%
  \raisebox{.9ex}{\scriptsize$\prime\prime$}\kern -0.1ex}}

\def\ion#1#2{#1$\;${\small\rm\@Roman{#2}}\relax}

\newcommand{\mmu}{\mbox{$<\!\!\mu\!\!>$}}

\title{AO Observations of Three Powerful Radio Galaxies}

\author{Wim de Vries\supit{a}, Wil van Breugel\supit{a}, and
Andreas Quirrenbach\supit{b}
\skiplinehalf
\supit{a}IGPP-LLNL, 7000 East Ave, Livermore, CA 94550, USA\\
\supit{b}Univ. of California at San Diego, San Diego, USA\\
}

\authorinfo{Email: devries1@llnl.gov, vanbreugel1@llnl.gov,
qui@cassir.ucsd.edu}

\begin{document} 
\maketitle 

\begin{abstract}

The host galaxies of powerful radio sources are ideal laboratories to
study active galactic nuclei (AGN).  The galaxies themselves are among
the most massive systems in the universe, and are believed to harbor
supermassive black holes (SMBH). If large galaxies are formed in a
hierarchical way by multiple merger events, radio galaxies at low
redshift represent the end-products of this process. However, it is not
clear why some of these massive ellipticals have associated radio
emission, while others do not. Both are thought to contain SMBHs, with
masses proportional to the total luminous mass in the bulge. It either
implies every SMBH has recurrent radio-loud phases, and the radio-quiet
galaxies happen to be in the ``low'' state, or that the radio galaxy
nuclei are physically different from radio-quiet ones, i.e. by having a
more massive SMBH for a given bulge mass. 

Here we present the first results from our adaptive optics imaging and
spectroscopy pilot program on three nearby powerful radio galaxies. 
Initiating a larger, more systematic AO survey of radio galaxies
(preferentially with Laser Guide Star equipped AO systems) has the
potential of furthering our understanding of the physical properties of
radio sources, their triggering, and their subsequent evolution.

\end{abstract}



\section{INTRODUCTION}
\label{sect:intro}  

Powerful radio galaxies provide a convenient way to investigate the
evolution of very massive galaxies over a large range in redshift.
Color properties of their host galaxies have been found to be
remarkably well represented by passively evolving stellar systems,
with typical masses of $5 - 10$ L$_\star$\cite{lilly88,vanbreugel98}.
Comparison with high redshift field galaxies\cite{cowie97} confirms
that radio galaxies form indeed the high-luminosity envelope. Inferred
formation redshifts for these systems have been as
high\cite{spinrad97} as $z_f > 10$, implying that by a redshift of 1
to 2, emission from these galaxies is dominated by an old ($>5$ Gyr)
stellar population. It is this homogeneity in population (and
emission) properties that makes the near-IR Hubble K-z relation so
tight.

Radio galaxy morphologies, when imaged at rest-frame optical
wavelengths, often show spectacular, clumpy structures aligned with
the radio source axes. This ``alignment effect'' appeared at odds with
the passive evolution inferred from the near-IR K-z diagram. Its exact
nature has remained unclear and evidence has been found for scattered
light from hidden quasar-like AGN, nebular re-combination continuum
and even jet-induced star formation\cite{mccarthy93}.

To investigate the morphological evolution of the stellar populations
of radio galaxies with redshift, it is therefore of interest to obtain
high spatial resolution at {\it infrared} wavelengths, where
AGN-related emission is fainter and the old stellar population
brighter. This effective isolation of just the stellar emission
component also provides a useful baseline against which we can
interpret the optical HST data. Again, given the rather uniform
stellar population, any color-deviations will stand out in a color
plot based on combining HST optical and Keck AO near-IR data
(cf. Fig.~\ref{colorMaps}). Features, such as dust-lanes, or compact
regions of enhanced starformation can be clearly detected against the
uniform backdrop of the underlying galaxy. Furthermore, the high
resolution near-IR images allows us to extract physically meaningful
luminosity profiles, which are much less affected by obscuring dust or
AGN related non-stellar emission. Using HST on a large sample of
(radio quiet) galaxies, Faber et al.\cite{faber97} found the shape of
the inner galaxy profile to correlate with various physical quantities
such as absolute luminosity and central velocity dispersion; provided
the profile can be fitted with a ``Nuker''
profile\cite{lauer95,byun96}. Up till now, only with HST was the
necessary spatial resolution attainable. With the Keck-AO system, this
can be improved upon by at least a factor of 4
(cf. Fig.~\ref{3c452nuker}). This allows us to intercompare more
distant radio galaxy hosts to ``normal'' ellipticals, which could
provide additional clues about why some galaxies are radio emitters
and others are not.

In principle high resolution spectroscopy can not only provide us with
diagnostic line ratios which constrain ionization mechanisms (i.e.,
starbursts vs. AGN\cite{hill99,vanzi98,black87}), and absorption
properties of circumnuclear features\cite{rudy99,thornton99,rhee00},
but it can also provide us with accurate measurements of nuclear
stellar velocity dispersions using the $^{12}$CO near-IR bandhead at
2.29$\mu$m restframe\cite{gaffney95,boeker99}. Given the very high
spatial (and spectroscopic) resolution of the observations, these will
provide nuclear black-hole (BH) mass estimates (analogous to B\"oker
\cite{boeker99} et al.). Values derived this way for our sample can
then be compared directly to recent results on the velocity---BH-mass
correlation for lower redshift galaxies\cite{gebhardt00,ferrarese00}.
However, a combination of instrumental throughput issues beyond
$\sim2.4$ $\mu$m, the anisoplanatism inherent to off-axis correction,
and just plain bad weather prevented us from achieving our planned
spectroscopic goals.

Either way, both the imaging and spectroscopy parts of the program
have yielded insights into the makeup of the stellar systems of the 3
powerful radio galaxies and their place in the elliptical galaxy
taxonomy. This illustrates the potential AO can offer us in
understanding the onset and subsequent evolution of powerful nuclear
radio sources, provided this pilot project is carried out on a much
larger sample and in a more complete fashion, ideally with a LGS
equipped AO system.

\section{Observations}

Our target galaxies have been selected from cross-correlations between
the NVSS radio catalog, the HST guide star catalog, and the HST
archive catalog. Positive hits have been checked against the USNO-A2.0
star catalog, as a fair fraction of the HST guide star catalog
consists of non-stellar objects (e.g., galactic nuclei), and we want
to make sure there is an AO star present.  Furthermore, we have
limited the redshift range to $0.015 < z < 0.10$ in order for the
$^{12}$CO spectroscopic band-head at 2.293 $\mu$m not to be redshifted
beyond 2.5 $\mu$m. The decreasing instrumental response and increase
in the thermal background (part of which is instrumental, too) are not
well suited for observing beyond this wavelength with the KECK NIRSPEC
instrument.  The sample and observations are listed in
Table~\ref{obsRes}.

As for the imaging part of the program, the high resolution H and
K-band AO images surpass the complementary R-band HST images in
resolution. In addition, the combination of the optical and near-IR
images (both with $<0.1$ arcsecond resolution) provides a powerful
tool for investigating the nuclear morphologies of these radio
galaxies in unprecedented detail (cf. Fig.~\ref{colorMaps}). The much
higher near-IR resolution of the Keck AO system compared to the HST
NICMOS cameras makes this possible for the first time.

Of the three radio sources (3C~403, 3C~405, and 3C~452), only 3C~452
was specifically imaged with NIRSPEC's slitviewing camera (SCAM). For
this purpose the smallest of the slits (1.3$\times$0.013\arcsec) was
inserted to minimize the light lost into this slit (and to SCAM), plus
the object was dithered across the field of view to further reduce the
impact of the slit. The other two sources were mainly spectroscopic
targets, and as such the SCAM images suffer from the presence of the
``large'' 3.96$\times$0.072\arcsec\ slit which consistently covered
the nucleus (since we were interested in the nuclear {\it spectrum}).

The SCAM camera, a 256$^2$ HgCdTe NICMOS array, has both a low
readout-noise and a high-throughput, and provides with its
0.0172\arcsec\ pixel scale a 4.4 square arcseconds field of view (cf.
NIRSPEC/AO manual). The NIRSPEC spectrograph was operated in low-
resolution mode, and resulted in a $\sim12.7$\AA\ instrumental
resolution at 2.4 $\mu$m ($R\approx 1900$), equivalent to a $\sim 160$
km/s velocity resolution. This matches best the expected velocity
dispersion of $\sim500$km/s (FWHM) of our objects, while still
retaining adequate S/N per resolution element. The high resolution
mode of the spectrograph ($R\approx 20 000$) would not.

The superb spatial resolution of the AO-NIRSPEC spectrograph allows
for an effective isolation of the spectral features related to the AGN
from the underlying galaxy. This, in principle, allows for accurate
BH-mass assessments (cf. Sect~\ref{spec}).

\begin{table}[b]
\begin{center}
\caption{Radio Source Sample + Observational Results\label{obsRes}}
\begin{tabular}{lllrr}
\\
Source Name & Coordinates (J2000)       & Redshift  &  GS Mag. & Separation \\
\hline
3C 403      & 19 52 15.39 $+$02 30 28.2 & 0.0590  & 10.7     &  8\arcsec  \\
3C 405      & 19 59 25.86 $+$40 43 50.9 & 0.05608 & 12.4     & 31\arcsec  \\
3C 452      & 22 45 48.85 $+$39 41 15.2 & 0.0811  & 10.8     & 13\arcsec  \\
\\
\end{tabular}
\begin{tabular}{lll}
\\
{\bf Imaging Observations} & Filter & \\
\hline
3C 452       & H-band & 5$\times$300s exposure, 24 Jun 2000\\
             &        & H = 13.5$\pm$0.2 \\
3C 452       & K$^\prime$-band & 5$\times$300s exposure, 24 Jun 2000\\
             &        & K$^\prime$ = 13.4$\pm$0.2 \\
\\
{\bf Spectroscopic Observations} & Filter & \\
\hline
3C 403       & NIRSPEC7 &  4$\times$900s exposure, 09 May 2001\\
             &          &  3.96$\times$0.072\arcsec slit\\
3C 405       & NIRSPEC7 &  8$\times$900s exposure, 26 Aug 2001\\
             &          &  3.96$\times$0.072\arcsec slit\\
3C 452       & NIRSPEC7 &  3$\times$1400s exposure, 31 Aug 2001\\
             &          &  3.96$\times$0.072\arcsec slit\\
\end{tabular}
\end{center}
\end{table}

\section{Imaging of 3C 452 - A Case Study}

The performance of the AO-system declines with distance from the
optical axis, but does so rather gracefully. The separation between
3C~452 and the guide star is $\sim11$\arcsec\, which is well inside
the isoplanatic angle of (up to) 45\arcsec\ radius for the Keck
system. Based on our observations on 3C 294 earlier in the
night\cite{quirrenbach01}, and experiences by other
observers\cite{larkin00}, we estimate that the Strehl-ratio for our
images declines from the $\sim0.2$ for the on-axis case of the guide
star, to $\sim0.1$ for the radio galaxy itself. Given the rather small
point-source component in the galaxy (cf. Sect.~\ref{AGNdecomp}) a
direct assessment of the Strehl-ratio is not possible. The FWHM of the
core in the K$^\prime$-band image is 55 milliarcseconds however, close
to the theoretical resolution limit of a 10m telescope (45 mas at
K$^\prime$).

Since our source is well resolved and effectively without an
unresolved component, the image resolution is for all practical
purposes comparable to the diffraction limit. In other words, even
though the Strehl-ratios are only about of 10\%\ of the values of a
perfect telescope, the resulting PSF is peaked enough compared to the
galaxy profile that its convolution does not significantly degrades
the input image. Inversely, since the convolved image is the one we
actually obtain, a deconvolution with either a model of actual PSF
does not significantly change the image properties. Image profile
parameters (cf. Sect.~\ref{profAnalysis}) between the unconvolved and
Richardson-Lucy deconvolved images were identical within their
errorbars.

\subsection{Morphological Parameters}

The H and K$^\prime$-band AO images have a resolution comparable to
HST WFPC2 optical images, making it for the first time possible to
construct color images which are not affected by resolution effects
(like WFPC2 - NICMOS color maps). 

\begin{figure}[t]
\begin{center}
\plottwo{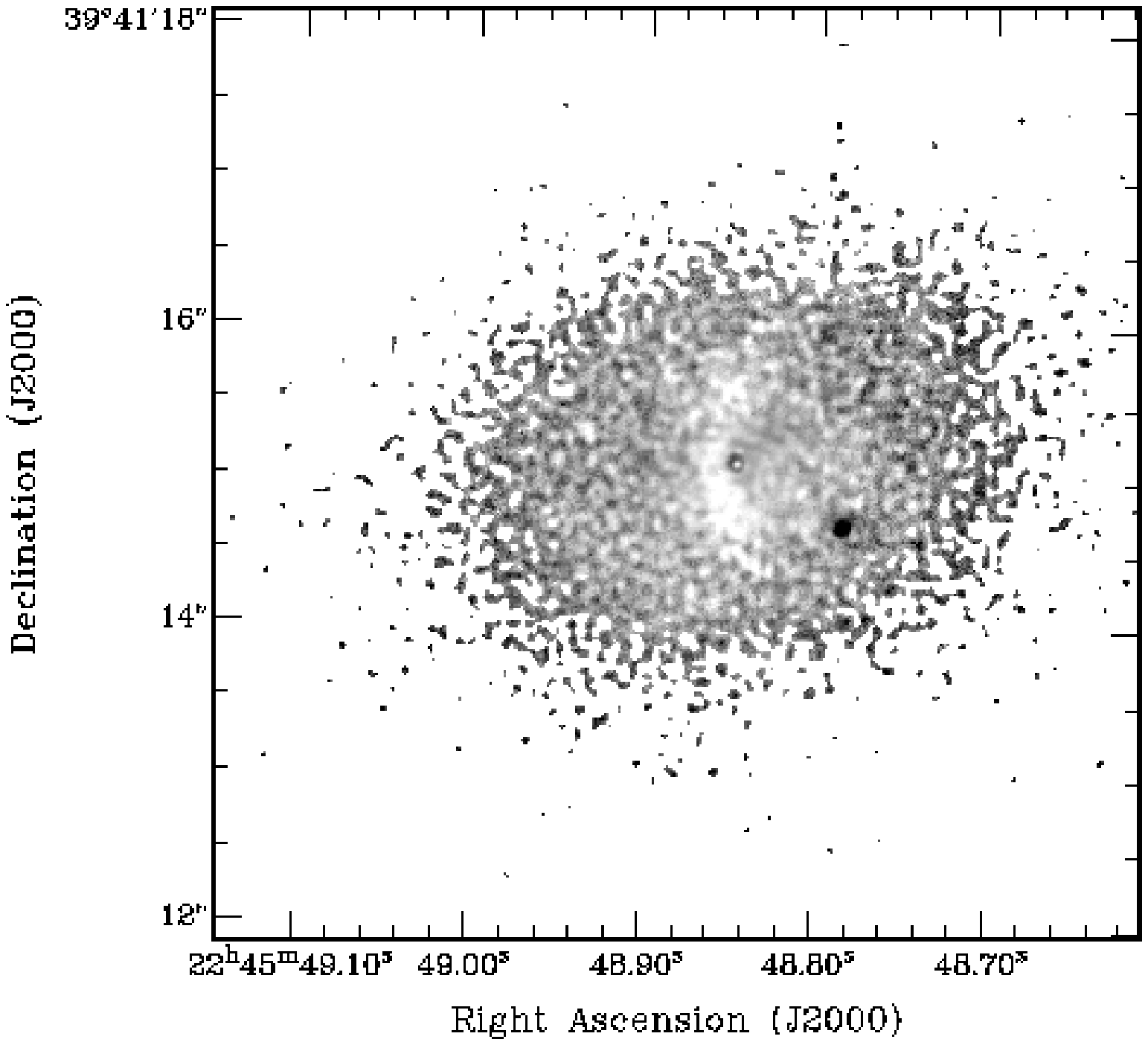}{0.462}{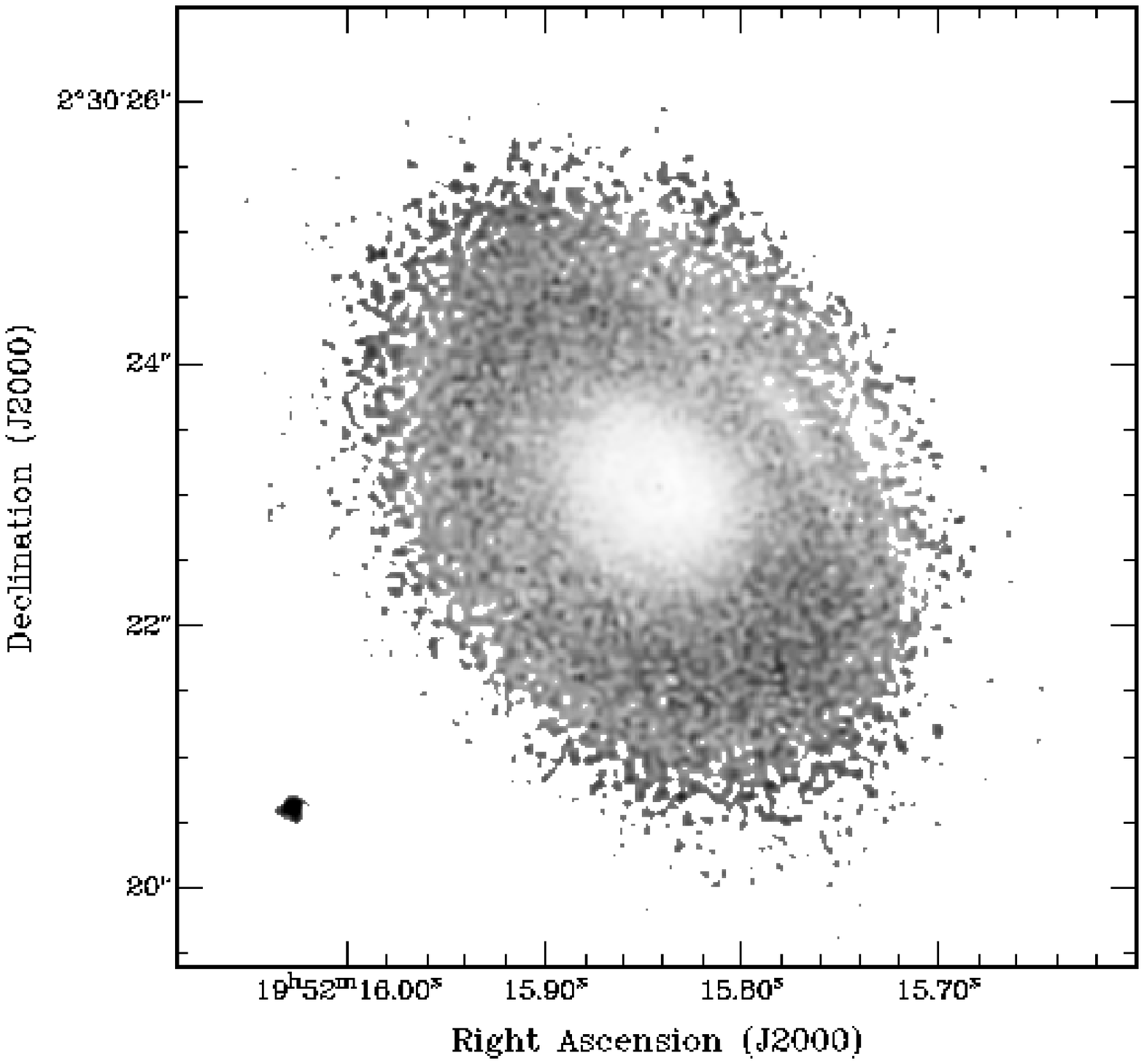}{0.45}
\caption{[Left Panel] Color Map of 3C 452. Both the HST WFPC2 image
and the KECK-AO H-band image have been renormalized to the same energy
flux levels (in units of ergs/s/cm$^2$/count/pixel).  The color coding
is as follows: Grey = neutral, White = more energy in H-band, Black =
more energy in R-band. Note the prominent dust lane, and the ``blue''
companion. [Right Panel] Same for 3C 403, with the key distinction
that the near-IR data has been modeled, based on actual data. The
color of the companion in the lower left corner is undefined: it was
not covered by this model. \label{colorMaps}}
\end{center}
\end{figure}

Fluxes in each of the images are converted into Watts ($\nu F_\nu$)
before dividing both images. This way the ratio represents a real
fractional energy excess.  The color maps are plots of the following
function:

\begin{equation} \label{chap8eqn1}
\mbox{ColorMap} = \frac{1 + f \times \mbox{CR}_{F702W} \mbox{[DN/s]}}
 {1 + \mbox{CR}_{H}\mbox{[DN/s]}} - 1
\end{equation}

\noindent where the factor $f$ is calculated by normalizing F702W and
H-band fluxes to the same energy level, after resampling the F702W
pixels onto the smaller H-band pixels (7 H-band pixels per WFPC2
pixel, area-wise). The factor works out to be: $f\approx50$.  The
constant 1 has been chosen in such a way as to yield approximately the
same standard deviation in the sky as in the original images. This
also suppresses large color variations in the noise dominated areas of
the map.  The colormaps are presented in Fig.~\ref{colorMaps}. For
comparison we included a modified color-map for the source 3C
403. Since the slit was so much more prominent in this source, we
actually used a model of the near-IR luminosity distribution instead
of the actual image to construct the color-map.

The smooth near-IR background luminosity distribution of 3C 452
provides a very good background to offset the optical HST data
against. This is quite dramatically demonstrated by the dust-lane,
which was hitherto only hinted at. The near-IR to optical color
baseline allows us to accurately assess the obscuring properties of
this dust lane. Following De Koff\cite{dekoff00} et al., we can infer
a dust mass for this torus from:

\begin{equation}
M_{dust}=\frac{\sum <A_\lambda> \mbox{[mag][kpc$^2$]}}
  {6\times10^{-6} \mbox{[mag][kpc$^2$][M$_\odot$]}},
\end{equation}

\noindent with the summation over the spatial extent of the obscuring
material, and $<A_\lambda>$ the mean absorption in magnitudes. Based
on this, we arrive at a $\sim1\times10^4$M$_\odot$ mass estimate for
the dust-lane in 3C 452. This is about an order of magnitude less (at
the same redshift) than dust-masses inferred for other 3C radio
galaxies\footnote{The dust-mass for 3C 452 listed in Table~2 of De
Koff et al. is wrong.}, illustrating the usefulness of AO observations
for morphological studies of the direct AGN environment.

\subsection{Profile Analysis}\label{profAnalysis}

\begin{figure}[t]
\begin{center}
\plottwo{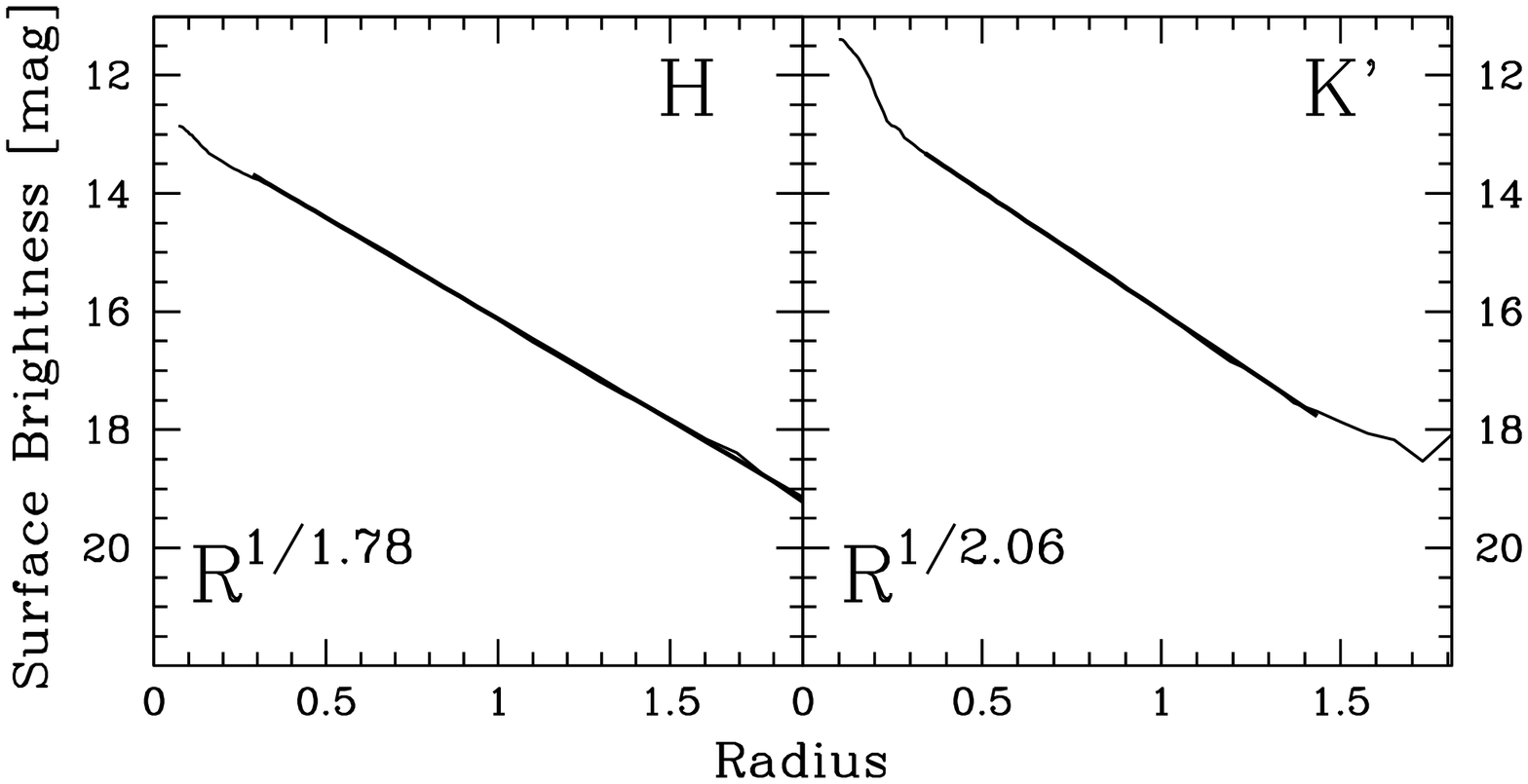}{0.645}{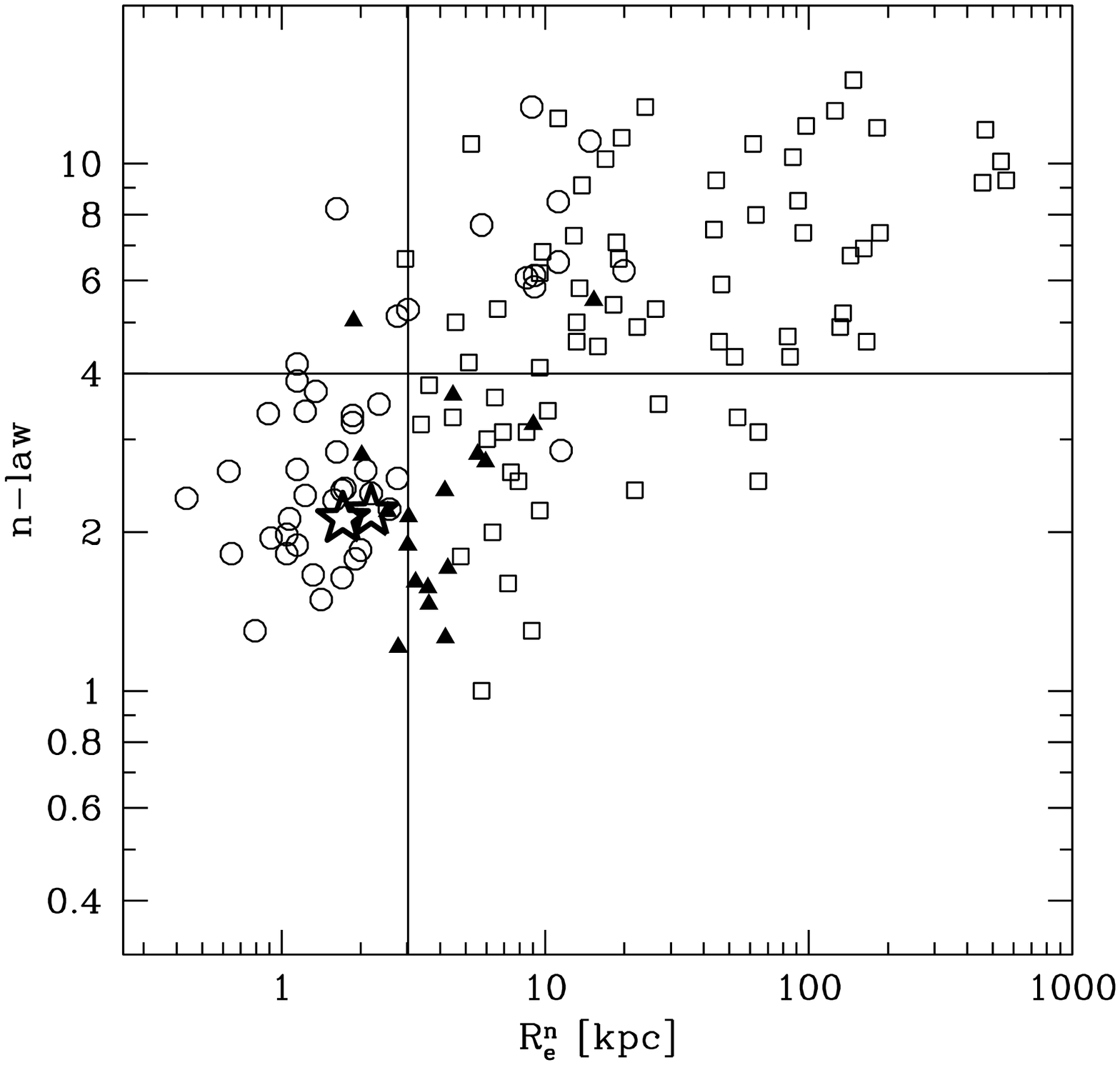}{0.342}
\caption{[Left Panel] Near-IR AO radial luminosity profiles of 3C~452,
with their best fitting $R^{1/n}$ law overplotted.  In the
K$^\prime$-band an unresolved nuclear component can be seen, something
not visible in the H-band. [Right Panel] Plot of Effective Radius
versus n-law index. Small ellipticals are in the lower left, large
Brightest Cluster Galaxies to the upper right corner. The solid
triangles are 3C radio galaxies (profiles measured with
NICMOS\cite{devries00}), and our AO source 3C~452 is indicated by the
two open star symbols (H and K$^\prime$-bands). \label{3c452nlaw}}
\end{center}
\end{figure}

\begin{table}[b]
\begin{center}
\caption{3C~452 Profile Results\label{profileData}}
\begin{tabular}{ll}
\\
{\bf n-law Fit}\\
\hline
H-band & $n=1.71\pm0.09$   R$_{\mbox{eff}}=2.12\pm0.01$ kpc \\
K$^\prime$-band & $n=2.19\pm0.04$   R$_{\mbox{eff}}=2.18\pm0.02$ kpc \\
\\
{\bf Nuker law Fit}\\
\hline
H-band & $\alpha=0.94$, $\beta=4.65$, $\gamma=0.12$, R$_{\mbox{break}}=1.68$ kpc \\
\end{tabular}
\end{center}
\end{table}

Luminosity profiles and their derived quantities -- effective radius
($R_e$), effective surface brightness ($\mu_e$), mean surface
brightness (\mmu), and their general shape are all important
quantities for discriminating between various objects. Spiral galaxies
usually have exponential profiles, whereas ellipticals are better
fitted with de Vaucouleurs' type laws ($\mu(r) \propto
R^{1/n}$). While the morphological differences between spirals and
ellipticals are obvious and ellipticals as a class appear rather
similar, they do exhibit significant differences in their profiles.
For instance, Faber\cite{faber97} et al. find a correlation between
absolute luminosity and profile shape, in the sense that the more
massive systems have shallower inner profiles. Less luminous
ellipticals have profiles which lack this ``core'', and remain steep
all the way up the resolution limit.  The physical size scale of this
cusp has been found to be on the order of 500
pc\cite{lauer95,faber97}.  This ``break radius'' corresponds to about
$0.35$\arcsec\ for 3C~452, or $\sim20$ pixels, enough for a detailed
fit. Again, this would not have been possible at lower resolutions.
First, we will fit the profile with a generalized de Vaucouleurs' law
($R^{1/n}$). This exponent $n$ also correlates with elliptical
type. Following Graham\cite{graham96} et al., we define:

\begin{equation} \label{genvauc}
\mu(r) = \mu_0 + \frac{2.5 b_n}{\ln(10)} \left(\frac{r}{r_e}\right)^{1/n}
\end{equation}

\noindent with $r_e$ the scale radius, $\mu_0$ the central surface
brightness, and the constant $b_n$ approximated by: $b_n \approx 2n -
0.327$. Note that in the $n=1$ case the profile is exponential, and
better represents spirals. The major axis luminosity profile of 3C~452
was fitted with this law, and the $\chi^2$ minimized fitting values
are given in Table~\ref{profileData}. A literature sample of radio
quiet ellipticals of varying size\cite{caon93,graham96}, and powerful
radio sources\cite{devries00} is presented in Fig.~\ref{3c452nlaw},
right panel. Our AO data on 3C~452 indicate that this is an
intermediate sized elliptical, possibly towards the small end of the
radio galaxy distribution. There is no indication, based on the
profile, that it has undergone any major merger in its history: the
profile lacks any hint of flattening towards the center. This seems to
be a common trait among powerful radio galaxies\cite{devries00}.

\begin{figure}[t]
\begin{center}
\plottwo{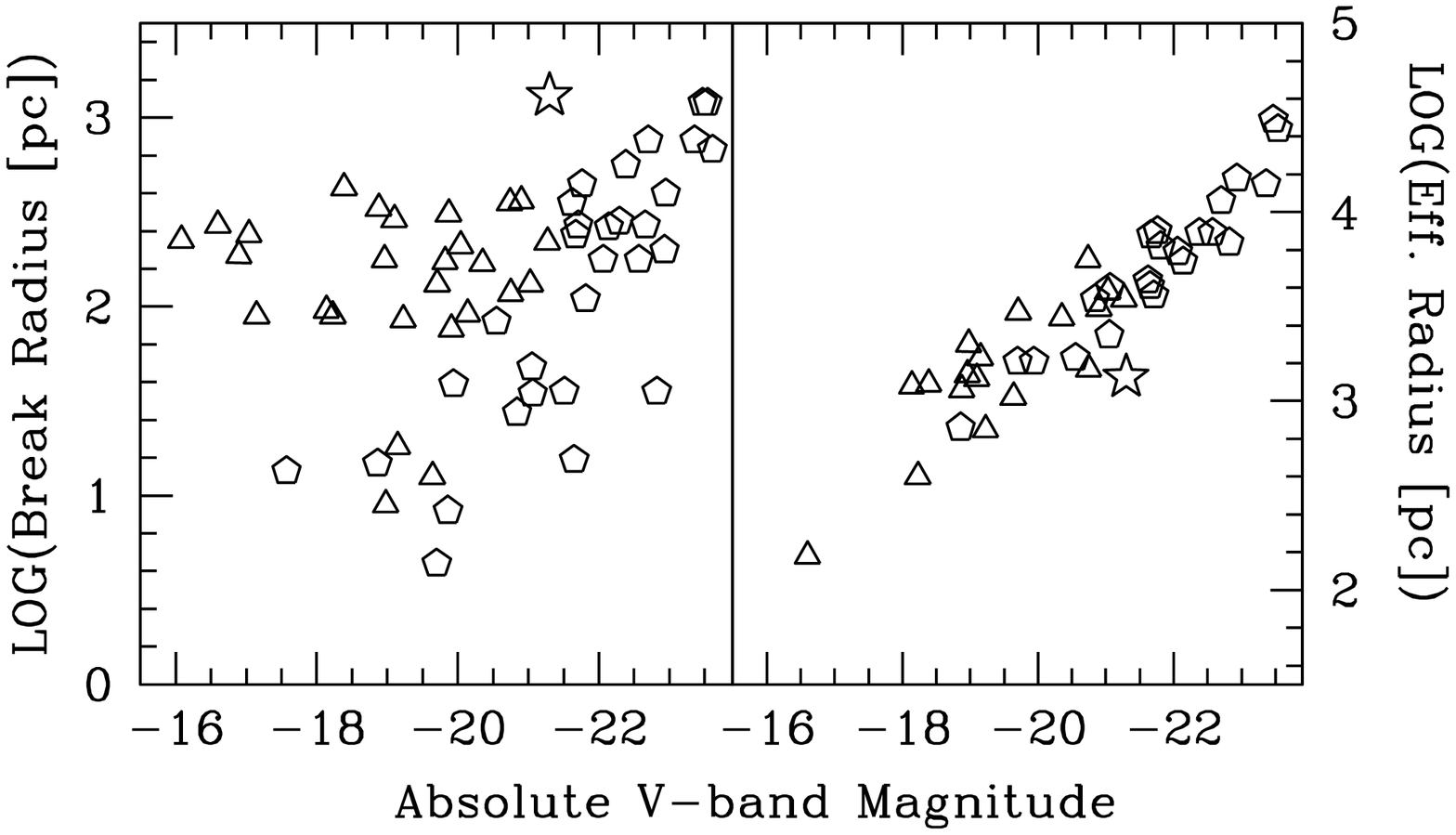}{0.62}{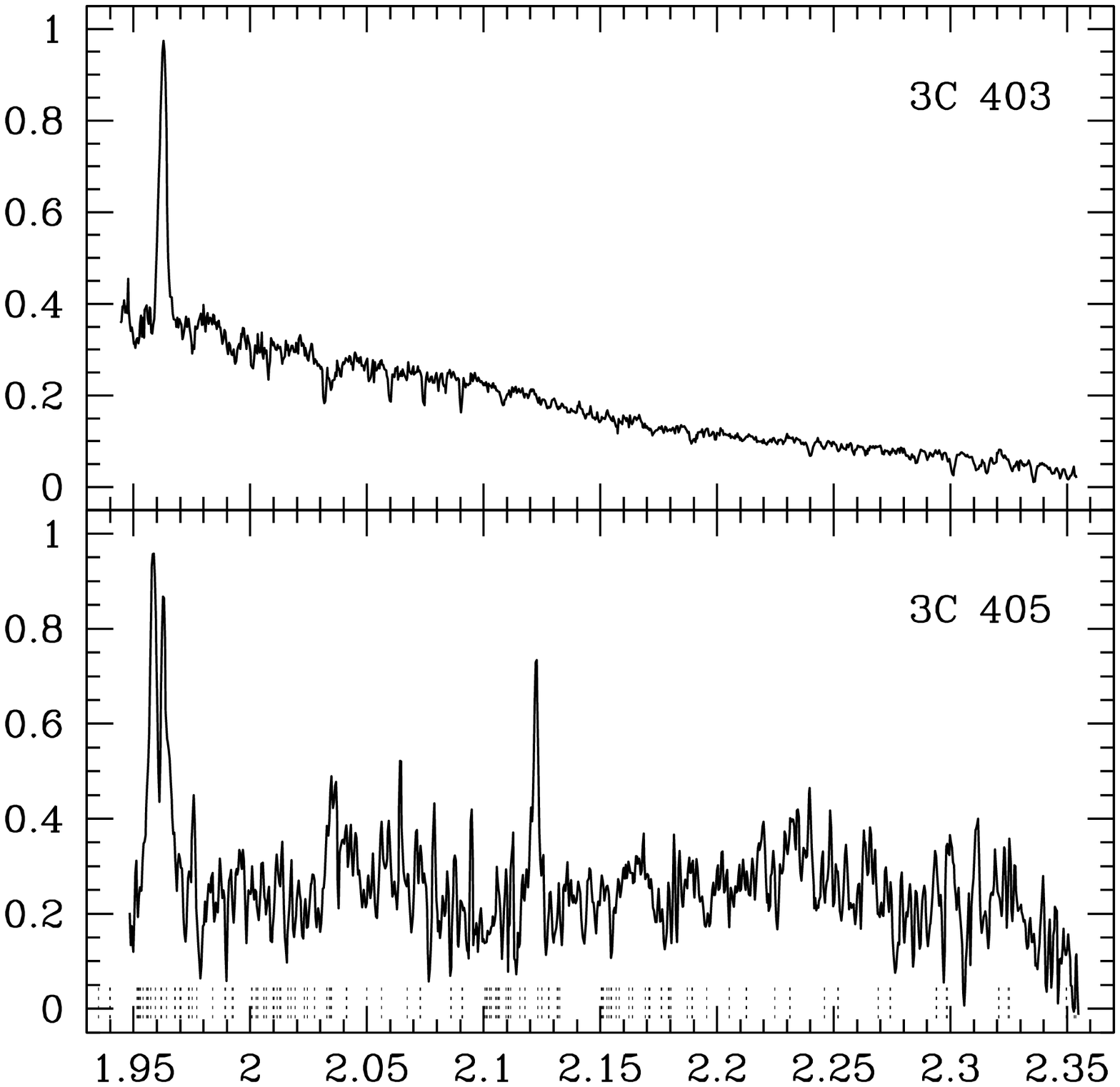}{0.37}
\caption{[Left Panel] 3C~452 (open star symbol) compared to
low-redshift (HST observed) ellipticals\cite{faber97}. Both the
Break-radius and the Effective radius correlate with absolute
luminosity. Open triangles: ``steep'' ellipticals ($\gamma<0.3$), open
pentagons: ``core'' ellipticals ($\gamma \ge 0.3$). [Right Panel]
Near-IR AO spectra of 3C~403 and 3C~405, plotted in the source
restframe with arbitrary flux units. Note the much higher S/N in the
spectrum of 3C~403. Lines are identified in Table~\ref{lines}. The
thin lines on the bottom of 3C~405's spectrum indicate the position of
atmospheric OH lines.
\label{3c452nuker}}
\end{center}
\end{figure}

The four free parameters of the Faber\cite{faber97} et al. profile law
(coined ``Nuker''-law) were fitted simultaneously using our $\chi^2$
minimizing simulated-annealing code. Results are listed in
Table~\ref{profileData}. Consistent with our n-law fit, the source is
classified as a ``steep'' source (indicated as open triangles in
Fig.~\ref{3c452nuker}), albeit with a brighter host galaxy. This may
be a selection effect though, since 3C~452 is at a much larger
distance than the literature galaxies. The main purpose of this
exercise is to demonstrate the feasibility of high resolution profile
analysis on these distant objects, enlarging the time baseline against
which to study evolutionary effects.

\subsection{AGN Luminosity Contribution}\label{AGNdecomp}

As can be seen in Fig.~\ref{3c452nlaw} (left panel), any deviation
from the smooth n-law profile will stand out. Of particular interest
is the nuclear (AGN) contribution to the total. By extrapolating the
n-law inwards, we can estimate the AGN contribution by measuring the
excess emission over the n-law. We actually performed a $\chi^2$
minimization of a variable pointsource contribution to the profile
shape, i.e., the goodness-of-fit was given by the least deviation from
a perfect n-law profile. The results are listed in
Table~\ref{profileData}. For our spectroscopy program, it was
necessary that the AGN would not contribute too much to the nuclear
emission, because it might wash out the stellar absorption signature
we were trying to measure. It is clear that with just a 4\%\ PSF
contribution in the K$^\prime$-band this condition is satisfied.

\section{Spectroscopy}\label{spec}

\begin{figure}[t]
\begin{center}
\plottwo{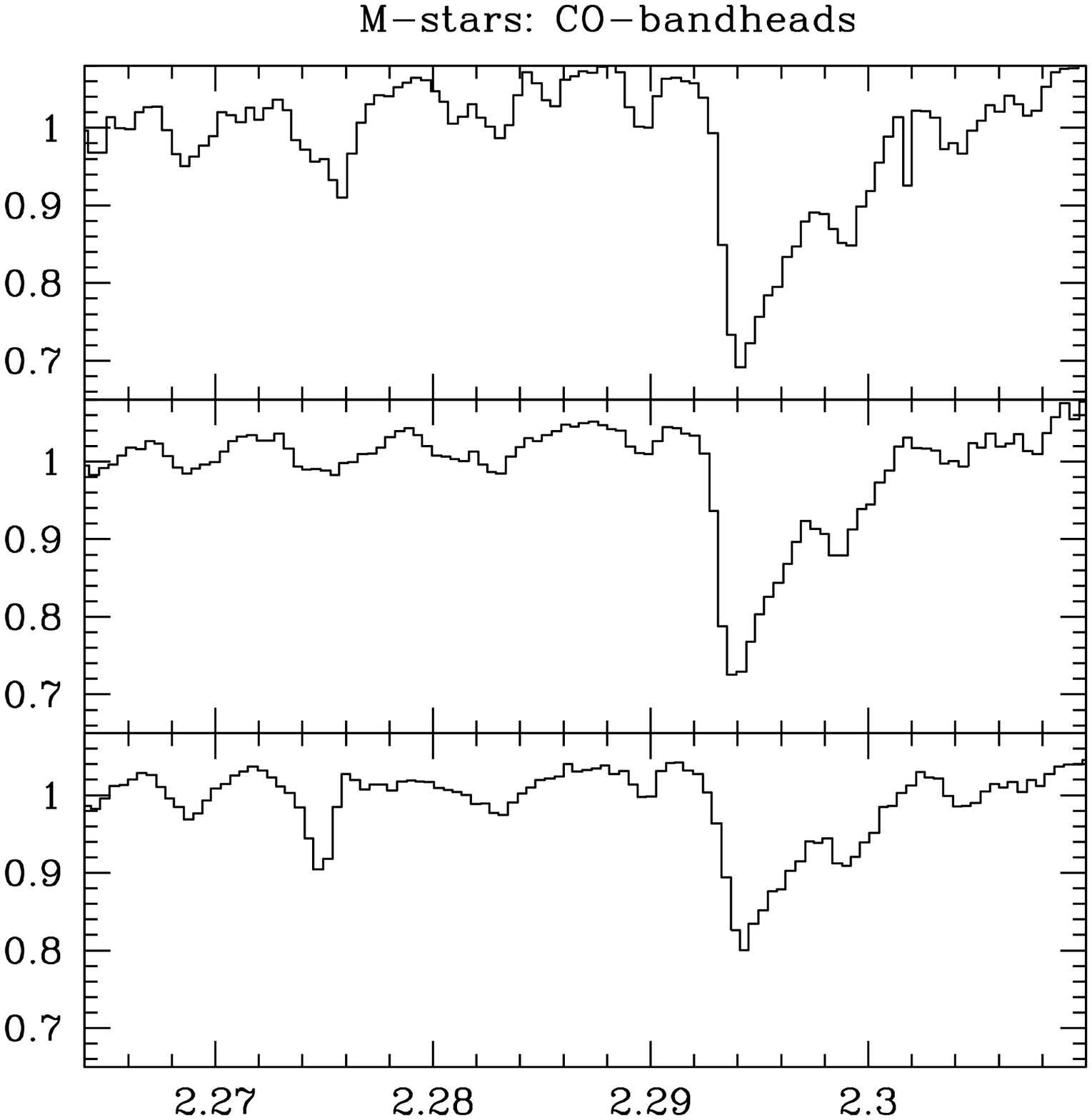}{0.35}{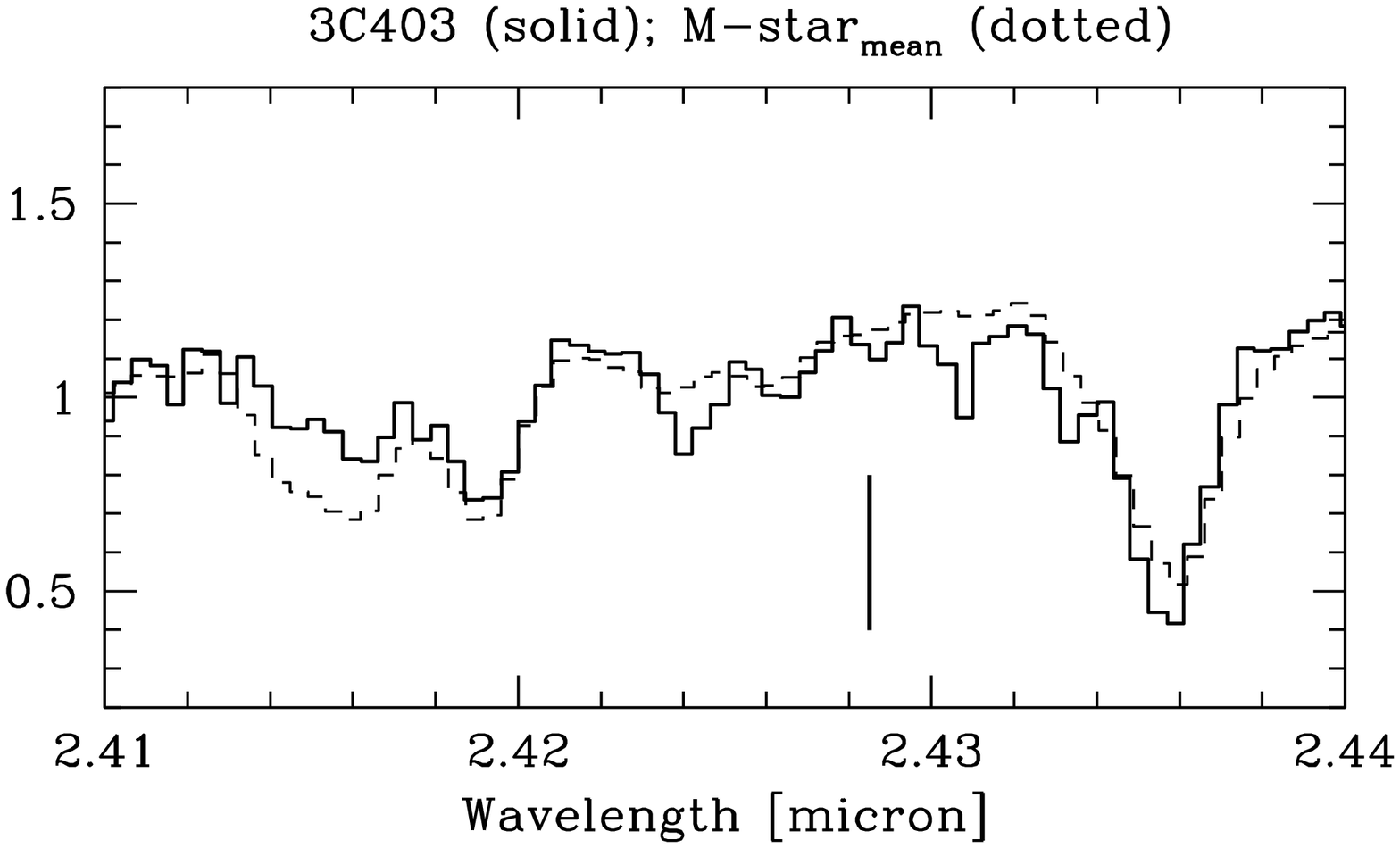}{0.63}
\caption{[Left Panel] $^{12}$CO bandheads as detected 
in our M-type template stars. [Right Panel] Actual
spectral region of 3C 403 (the radio source with the
highest S/N spectrum) which should contain the redshifted
bandhead (marked with the vertical line). A mean M-star
spectrum is overplotted. No hint of a bandhead is present
in 3C 403. \label{spectra}}
\end{center}
\end{figure}

The original idea was to combine the very high spatial resolution and
the near-IR capability of the NIRSPEC instrument to measure the exact
slope of the $^{12}$CO bandhead at 2.293$\mu$m restframe. This method
has been successfully applied before\cite{boeker99}, and since this
line is formed in outer envelopes of cool stars, it provides a clean
measure of the {\it stellar} velocity dispersion, and is not affected
by thermal broadening: the slope of the bandhead becomes more shallow
with an increase of the velocity dispersion of the stars.  With
NIRSPEC-AO and its complement of very small slits, we would be able to
obtain spectra of the region very close to the nucleus, a region
presumably gravitationally influenced by the central massive black
hole (the slit width of 0.072\arcsec\ corresponds to $\sim$100 pc at
the redshift of 3C~452).

Unfortunately, the spectroscopic data on 3C~405 and 3C~452 do not have
the required S/N. The one source with enough S/N (3C~403) does not
exhibit any sign of a CO bandhead at its correct (redshifted) position
(cf.  Fig.~\ref{spectra}, right panel). The instrumental setup clearly
was able to measure bandheads (left panel), so in the case of 3C~403
it might be central obscuration that hides this spectral signature.
This (more or less uniform) obscuration is hinted at in the colormap
of Fig.~\ref{colorMaps}, right panel.

\begin{table}[b]
\begin{center}
\caption{Near-IR emission lines\label{lines}}
\begin{tabular}{lllll}
\\
{\bf 3C~403} \\
Line & $\lambda$ & FWHM & $\sigma$ [km/s] & Redshift \\
\\
Br-$\delta$ 19451\AA\   & 20591\AA\ &          &     & 0.0586\\
$[$\ion{Si}{6}$]$ 19629\AA\ & 20777\AA\ & 35.4\AA\ & 218 & 0.0585\\
$[$\ion{Ca}{8}$]$ 23211\AA\ & 24569\AA\ &          &     & 0.0585\\
\hline
\\
{\bf 3C~405}\\
Line & $\lambda$ & FWHM & $\sigma$ [km/s] & Redshift \\
\\
H$_2$ 1-0S(3) 19576\AA\ & 20684\AA\ & 34.5\AA\ & 213 & 0.0566\\
$[$\ion{Si}{6}$]$ 19629\AA\ & 20746\AA\ & 35.3\AA\ & 217 & 0.0569\\
H$_2$ 1-0S(2) 20338\AA\ & 21489\AA\ &          &     & 0.0566\\
H$_2$ 1-0S(1) 21218\AA\ & 22417\AA\ & 19.4\AA\ & 110 & 0.0565\\
\hline
\end{tabular}
\begin{tabular}{p{9.5cm}}
{\sc Note --} FWHM's have been corrected for the instrumental 
resolution of 12.7\AA.
\end{tabular}
\end{center}
\end{table}

\subsection{Emission Lines}

Though not strictly needing AO, we did measure several emission lines
in 3C~403 and 3C~405 (3C~452 did not have any, at least not to the
level detected). The measurements are listed in Table~\ref{lines}. The
high excitation line $[$\ion{Si}{6}$]$, due to either UV irradiation
by the central AGN, or to powerful shocks close to the AGN, is present
in both spectra, with almost identical widths. It should be noted that
these widths (as $\sigma$ in km/s) are very compatible with stellar
velocity dispersions as measured for radio galaxies\cite{bettoni01},
but not necessarily due to gravitational motion.

\section{Summary}

We demonstrated the potential AO observations have to radio galaxy
host studies. It equals, and complements, HST's optical imaging in
terms of resolution, extending the realm of detailed nuclear
environment imaging into the near-IR. The spectroscopic setup we used
(NIRSPEC + KECK) was not ideally suited for our project, however.
When working close to the diffraction limit, even 10m class telescopes
are photon starved.

Based on this pilot study, AO imaging (and perhaps AO spectroscopy) on
a large, and systematically set up sample can significantly further
our knowledge of these large galaxies and their massive black holes.

\acknowledgments

This work has been supported in part by the National Science
Foundation Science and Technology Center for Adaptive Optics, managed
by the University of California at Santa Cruz under cooperative
agreement No.  AST-98-76783.  WDV and WVB's work was performed under
the auspices of the U.S. Department of Energy, National Nuclear
Security Administration by the University of California, Lawrence
Livermore National Laboratory under contract No. W-7405-Eng-48. Part
of this work was funded by NASA HST grant GO8183.

\bibliography{spiePaperAPH}   
\bibliographystyle{spiebib}   

\end{document}